\documentclass[twocolumn]{aastex701}

\usepackage{gensymb}
\usepackage{amsmath}

\begin{document}

\title{Wind power as a technosignature on M-dwarf planets}

\correspondingauthor{Jacob Haqq-Misra}

\author[0000-0003-4346-2611]{Jacob Haqq-Misra}
\affiliation{Blue Marble Space Institute of Science, 600 1st Ave, Floor 1, Seattle, WA, 98104, USA}
\email[show]{jacob@bmsis.org}

\author{Mykhaylo Danylov}
\affiliation{90 Sierra Vista, Apt \#127, Mountain View, CA, 94043, USA}
\email[]{mikhaildanilov@myyahoo.com}

\begin{abstract}
We suggest that the large-scale deployment of wind turbines on an M-dwarf planet could produce observable technosignatures. Motivated by observations of hypersonic wind velocities on WASP-127\,b, we note that the atmospheres of such planets could serve as vast reservoirs of energy for an extraterrestrial civilization. A large-scale deployment of wind turbines in a hypersonic environment would produce heated shock waves in the hypersonic stream, cause strong frictional heating from the rotation of the blades, and be a source of infrared radiation. We mention possible scenarios that could lead to the deployment of wind turbines on a gas giant and also note that similar features could exist on terrestrial M-dwarf planets. The idea that aerodynamic peculiarities could be a technosignature is worth keeping in mind as ground- and space-based exoplanet observations continue to improve.
\end{abstract}



\section{Introduction}
\label{sec:introduction}

The search for extraterrestrial life seeks to detect remotely detectable evidence of life, known as \textit{biosignatures}. Global-scale technology on Earth is a product of biology, which suggests possible \textit{technosignatures} that---if observed---would be compelling evidence for the presence of extraterrestrial technology. The search for technosignatures today includes ongoing surveys of nearby stars with radio and optical telescopes \cite[e.g.,][]{price2020breakthrough}. Others have analyzed stellar databases for anomalous infrared emission that would be associated with megastructures, such as Dyson spheres \citep[e.g.,][]{suazo2024project}. Studies are also exploring the potential for ground- and space-based telescopes to detect spectral features in exoplanetary atmospheres that would indicate the presence of pollutants \citep[e.g.,][]{lustig2023earth}. These and other approaches \citep[see, e.g.,][for reviews]{socas2021concepts,haqq2022searching} all seek to use the tools of astronomy to search for anomalies that may be associated with extraterrestrial technology. 

In this note, we discuss the idea that a large-scale deployment of wind turbines on an M-dwarf planet could produce observable technosignatures.

\section{Hypersonic Anomalies}

Planets in synchronous rotation around M-dwarf stars generate strong atmospheric winds, due in part to the strong thermal gradient between the substellar hemisphere in perpetual day and anti-stellar hemisphere in perpetual night. Observations of WASP-127\,b with the European Southern Observatory's Very Large Telescope have detected a double-peaked signal of H$_2$O and CO that can be explained by the presence of a hypersonic jet, with equatorial winds of 9\,km\,s$^{-1}$ (greater than the re-entry velocity of spacecraft on Earth) along with evidence of a strong thermal gradient across the terminator \citep{nortmann2025crires}. Such observations demonstrate the viability of using various spectral tracers in transiting exoplanets to infer wind velocities \citep[e.g.,][]{keles2021spectral}. 

Hypersonic winds could serve as a vast reservoir of energy for a technological civilization. The original idea envisioned by \citet{dyson1960search} was that a future technological civilization (on Earth, or elsewhere) could accommodate its growth in energy consumption to harness the entire luminous output of the sun by constructing a spherical shell or swarm of solar-collecting elements. This would also produce infrared waste energy that could serve as a technosignature. But perhaps instead of space-based solar collectors, an energy-intensive civilization around an M-dwarf host star might deploy a large number of electrical wind turbines across the terminator of a planet with hypersonic winds in order to satisfy its energy needs. 

A large-scale deployment of wind turbines on a synchronously rotating M-dwarf planet would produce powerful heated shock waves in the hypersonic stream (from the outer protective components of each turbine), which could be discernible in observations. Strong friction within the boundary layer would also cause high temperatures up to 1000$\degree$C (and even more if approaching Mach 10) in the outer protective components of the turbines. A large deployment of turbines would also be a source of infrared radiation. Theoretical predictions for the behavior of shock waves at hypersonic speeds \citep[e.g.,][]{hayes2012physics}, as well as measurements of hypersonic vehicles on Earth, could help to identify aerodynamical and thermodynamical features that can uniquely identify large-scale technology. If such anomalous aerodynamic features were observed on an exoplanet and could not also be explained by other processes, then they would constitute a compelling technosignature. 

The specific features of any anomalous shock waves or jets resulting from a large-scale deployment of wind turbines would depend on planetary properties that include the mass, radius, atmospheric composition, and layered structure. Gas giants like WASP-127\,b would not have a terrestrial surface for affixing turbine technology, so any resulting aerodynamic features would depend on details such as the depth of placement of the turbines and their spatial distribution. Asymmetric features inherent to the planet itself (such as the Great Red Spot on Jupiter) could contribute to unique properties of any shock waves resulting from turbines. Further theoretical studies with numerical models would be required to make quantitative predictions of expected features for a given exoplanet, which would be particularly useful if any candidate planets are observed with hypersonic anomalies.

Large-scale deployment of wind turbines on a terrestrial M-dwarf planet could also cause perturbations to the planetary atmosphere. For these planets, topographical features like high mountains could contribute aerodynamic features in addition to any from technology. If the maximum jet velocity on a terrestrial M-dwarf planet is subsonic, then there may be fewer uniquely identifiable features that could distinguish the presence of technology. But if supersonic or hypersonic winds are possible on some terrestrial M-dwarf planets, then the same possibility could remain of finding shock waves, temperature anomalies, and infrared excesses that result from planetary-scale technology.

\section{Conclusion}

Gas giants like WASP-127\,b serve as examples of environments with hypersonic velocities that hypothetically could be a source of energy harnessed by extraterrestrial technology. But this is not to say that gas giants are inhabited; indeed, if life is possible on gas giants like WASP-127\,b, then it almost certainly bears little resemblance to any life on Earth. The high-kinetic energy environment of such planets would not be conducive to life as we know it, with threats to biology even at velocities as low as Mach 2. We cannot preclude the possibility of unusual forms of life arising on such planets, perhaps even developing to the point of being able to construct technology; however, we also cannot easily predict evolutionary trajectories in such exotic environments. Another possibility is that life and technology first arise elsewhere on a terrestrial planet and then later seek to harness the energy on an uninhabited gas giant with hypersonic winds. This would be a case of a planet as a ``service world'' \citep[][]{wright2022case}, utilized for the deployment of planetary-scale technology in order to support the needs of a civilization that lives elsewhere. 

In summary, we suggest that the large-scale deployment of wind turbines on an M-dwarf exoplanet could lead to observable technosignatures, particularly if on a planet with hypersonic winds. The associated technosignatures would be anomalous shock waves, anomalous heating from the friction of the rotating blades, and infrared excesses observed along or near the planet's terminator. We do not ascribe any likelihood or occurrence rate for such extraterrestrial technology, but we do suggest that it is worth noting the possibility of such technosignatures as exoplanet observations continue to improve.

\begin{acknowledgments}
JHM gratefully acknowledges support from the NASA Exobiology program under grant 80NSSC22K1009.
\end{acknowledgments}

\begin{contribution}
JHM was responsible for writing and submitting the manuscript. MD came up with the initial research concept and edited the manuscript.
\end{contribution}

\newpage

\bibliography{main}{}
\bibliographystyle{aasjournalv7}



\end{document}